\documentclass[preprintnumbers, prd, showpacs, floatfix,
onecolumn, preprintnumbers, letterpaper,
superscriptaddress,nofootinbib]{revtex4}
\usepackage{amsfonts}
\usepackage{amsmath}
\usepackage{amssymb,epsf}
\usepackage{latexsym}
\usepackage{graphicx,epsfig}
\usepackage{amssymb}
\usepackage{subfigure}
\usepackage[colorlinks=true,citecolor=blue,linkcolor=blue,urlcolor=black]{hyperref}
\usepackage{epstopdf}

\begin{document}

\title{Holographic Conductivity for Logarithmic Charged Dilaton-Lifshitz
Solutions}
\author{A. Dehyadegari}
\affiliation{Physics Department and Biruni Observatory, Shiraz University, Shiraz 71454,
Iran}
\author{A. Sheykhi}
\email{asheykhi@shirazu.ac.ir}
\affiliation{Physics Department and Biruni Observatory, Shiraz University, Shiraz 71454,
Iran}
\affiliation{Research Institute for Astronomy and Astrophysics of Maragha (RIAAM), P.O.
Box 55134-441, Maragha, Iran}
\author{M. Kord Zangeneh}
\email{mkzangeneh@shirazu.ac.ir}
\affiliation{Physics Department and Biruni Observatory, Shiraz University, Shiraz 71454,
Iran}

\begin{abstract}
We disclose the effects of the logarithmic nonlinear electrodynamics on the
holographic conductivity of Lifshitz dilaton black holes/branes. We analyze
thermodynamics of these solutions as a necessary requirement for applying
gauge/gravity duality, by calculating conserved and thermodynamic quantities
such as the temperature, entropy, electric potential and mass of the black
holes/branes. We calculate the holographic conductivity for a $(2+1)$%
-dimensional brane boundary and study its behavior in terms of the frequency
per temperature. Interestingly enough, we find out that, in contrast to the
Lifshitz-Maxwell-dilaton black branes which has conductivity for all $z$,
here in the presence of nonlinear gauge field, the holographic conductivity
do exist provided $z\leq3$ and vanishes for $z>3$. It is shown that
independent of the nonlinear parameter $\beta$, the real part of the
conductivity is the same for a specific value of frequency per temperature
in both AdS and Lifshitz cases. Besides, the behavior of real part of
conductivity for large frequencies has a positive slope with respect to
large frequencies for a system with Lifshitz symmetry whereas it tends to a
constant for a system with AdS symmetry. This behavior may be interpreted as
existence of an additional charge carrier rather than the AdS case, and is
due to the presence of the scalar dilaton field in model. Similar behavior
for optical conductivity of single-layer graphene induced by mild oxygen
plasma exposure has been reported.
\end{abstract}

\maketitle

\section{INTRODUCTION}

The idea of correspondence between gravity in an anti-de Sitter (AdS) spaces
and a conformal field theory (CFT) living on its boundary (AdS/CFT) \cite%
{Mal,Wit,gub} have been successful in many theories like superconductors,
quark-gluon plasma and entanglement entropy. In recent years, the extension
of AdS/CFT correspondence to other gauge field theories and various
spacetimes (gravity theories) have got a lot of enthusiasm and usually
called gauge/gravity duality in the literatures. It has been well
established that the gauge/gravity duality provides powerful tools for
exploring dynamics of strongly coupled field theories and physics of our
real Universe. Recently, an interesting application of gauge/gravity duality
in condensed matter physics was suggested by Hartnoll, et. al. \cite%
{Har1,Har2} who demonstrated that some properties of strongly coupled
superconductors have dual gravitational descriptions. Such strongly coupled
superconducting phases of the boundary field theory are termed \textit{%
holographic superconductors} in the literatures.

On the other hand, the dynamics of many condensed matter systems near the
critical point can be described by a relativistic CFT or a more subtle
scaling theory respecting the Lifshitz symmetry \cite{Lif}%
\begin{equation}
t\rightarrow \lambda ^{z}t,\text{ \ \ \ \ }\vec{\mathbf{x}}\rightarrow
\lambda \vec{\mathbf{x}}.
\end{equation}%
The spacetime which supports above symmetry on its $r$-infinity boundary is
known as Lifshitz spacetime and has the line element \cite{Lif} 
\begin{equation}
ds^{2}=-\frac{r^{2z}}{l^{2z}}dt^{2}+\frac{l^{2}dr^{2}}{r^{2}}+r^{2}d\vec{%
\mathbf{x}}^{2},  \label{lifmet}
\end{equation}%
where $z$ is dynamical critical exponent. Black hole spacetime with
asymptotic Lifshitz symmetry have been widely investigated in the
literature. For example, thermodynamics of asymptotic Lifshitz black
solutions in the presence of massive gauge fields has been studied in \cite%
{Deh1}. The generalization to include the higher curvature corrections terms
to Einstein gravity and thermodynamics of asymptotically Lifshitz black hole
spacetimes were explored in \cite{hcc}. The studies were also extended to
dilaton gravity. In this regards, thermal behavior of uncharged \cite{peet}
and linearly charged \cite{tario} Lifshitz black holes in the context of
dilaton gravity have been explored. When the gauge field is in the form of
power-law Maxwell field, a new class of analytic topological Lifshitz black
holes with constant curvature horizon in four and higher dimensional
spacetime were constructed in \cite{Deh2}. A class of black brane solutions
of an effective supergravity action in the presence of a massless gauge
field, which contains Gauss-Bonnet term as well as a dilaton field, with
Lifshitz asymptotic have been investigated in \cite{Deh3}.

It is also interesting to study other physical properties of systems with
Lifshitz symmetry such as conductivity by applying the gauge/gravity duality 
\cite{pang,SupCondLif,CondLif,hollif,KDSD}. The holographic conductivity of
an Abelian Higgs model in a gravity background which is dual to a strongly
coupled system at a Lifshitz-fixed point, were explored in \cite{SupCondLif}%
. Other studies on the holographic superconductors with asymptotic Lifshitz
symmetry were carried out in \cite{1210,1211,1311,1402}. The behavior of
holographic conductivity for linearly charged Lifshitz black branes has been
studied in \cite{wu1,wu2} for $1\leq z\leq 2$. It is also of great
importance to investigate the effects of nonlinear electrodynamics on the
holographic conductivity. In \cite{KDSD}, the holographic conductivity for $%
4 $-dimensional Lifshitz black branes in the presence of nonlinear
exponential electrodynamics \cite{hendiexp} has been explored. The
pioneering study on the nonlinear electrodynamics was done by Born and
Infeld (BI) in $1934$ \cite{BI} who considered a Lagrangian of the form \cite%
{BI} 
\begin{equation}
L_{\mathrm{BI}}=4\beta ^{2}\left( 1-\sqrt{1+\frac{F^{2}}{2\beta ^{2}}}%
\right) ,  \label{LBI}
\end{equation}%
where $\beta $ is called the nonlinear parameter with dimension of mass, $%
F=F_{\mu \nu }F^{\mu \nu }$ and $F_{\mu \nu }=\partial _{\lbrack \mu }A_{\nu
]}$ in which $A_{\nu }$ is the gauge potential. It has been shown that BI
nonlinear electrodynamics is capable to remove the divergency of the
electric field of a point-like charged particle at its location as well as
the divergency of its self-energy. In addition to BI Lagrangian, other
BI-like nonlinear electrodynamics in the context of gravitational field have
been introduced. Among them, the so called logarithmic nonlinear, which
introduced by Soleng \cite{Soleng}, have got a lot of attention, in recent
years. The Lagrangian density of the logarithmic gauge field is given by 
\cite{Soleng} 
\begin{equation}
L\left( F\right) =-8\beta ^{2}\ln \left( 1+\frac{F}{8\beta ^{2}}\right) .
\end{equation}%
In the framework of dilaton gravity, thermal stability and thermodynamic
geometry of a class of black hole spacetimes in the presence of logarithmic
nonlinear electrodynamics have been explored in four \cite{SNZ1} and higher
dimensional spacetime \cite{SNZ2}. Also, a class of spinning magnetic
dilaton string solutions which produces a longitudinal nonlinear
electromagnetic field, in the presence of logarithmic nonlinear source, were
explored in \cite{SheyM}. These solutions have no curvature singularity and
no horizon, but have a conic geometry \cite{SheyM}. In this paper, we would
like to consider a class of asymptotically Lifshitz black hole/brane
solutions of Einstein-dilaton gravity in the presence of logarithmic
nonlinear electrodynamics and study the thermodynamics of them as a
necessary requirement for a system on which we intend to apply gauge/gravity
duality. We shall also calculate the holographic conductivity of linearly
and nonlinearly charged $4$-dimensional black brane solutions for all values
of $z$\ and disclose the effects of nonlinear gauge field on the
conductivity.

This paper is structured as follows. In the next section, we introduce the
action and construct a new class of asymptotic Lifshitz black hole/brane
solutions of Einstein-dilaton gravity in the presence of logarithmic
nonlinear electrodynamics. In section \ref{Therm}, we study thermodynamics
of the nonlinear Lifshitz black hole/brane solutions and calculate conserved
and thermodynamics quantities. We also verify the validity of the first law
of thermodynamics on the horizon. In section \ref{Cond}, we study the
holographic conductivity of two-dimensional systems for both linear Maxwell
and logarithmic nonlinear electrodynamics. We also plot the behavior of real
and imaginary parts of holographic conductivity for asymptotic AdS and
Lifshitz solutions. We finish our paper with concluding remarks in the last
section.

\section{Action and Lifshitz solutions\label{lifsol}}

\begin{figure*}[t]
\centering{%
\subfigure[$k=0$, $n=4$]{
   \label{fig1a}\includegraphics[width=.46\textwidth]{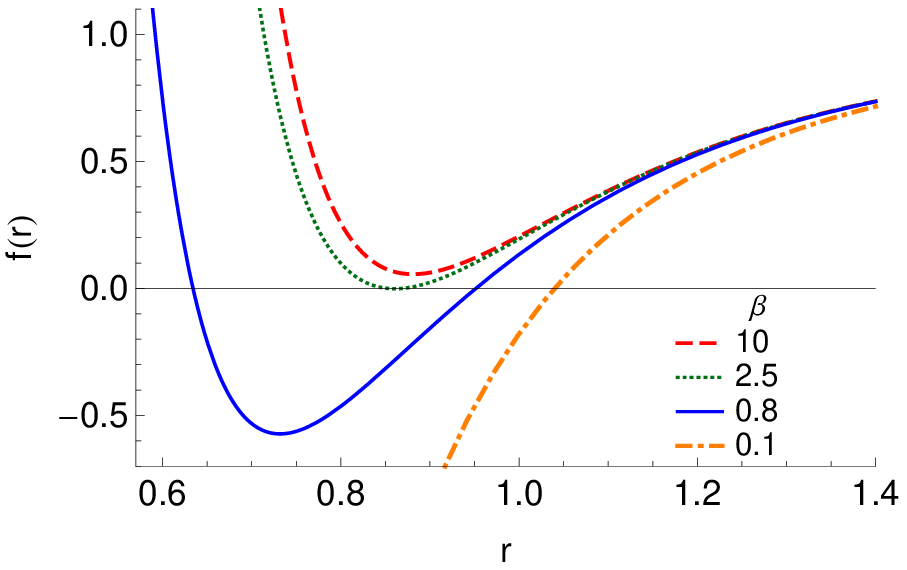}\qquad}} 
\subfigure[$k=1$, $n=5$]{
   \label{fig1b}\includegraphics[width=.46\textwidth]{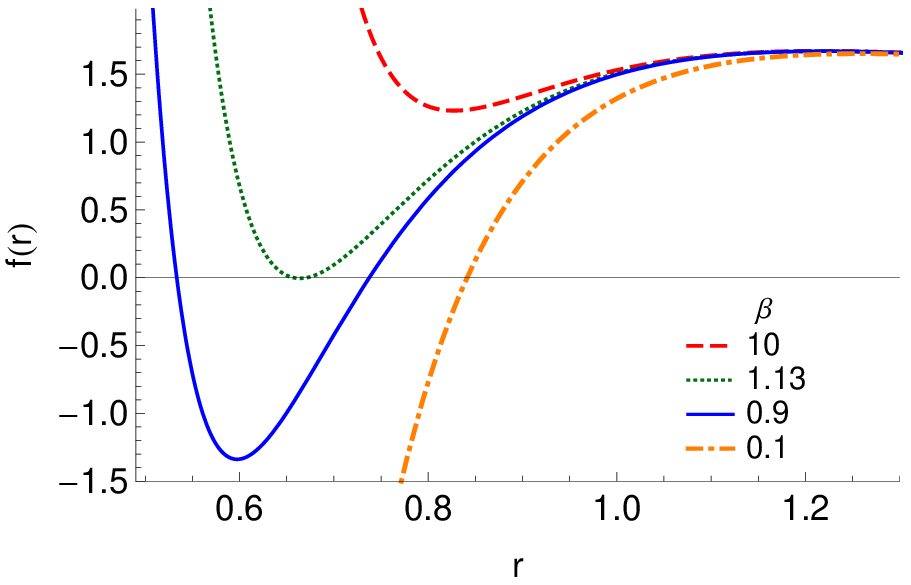}}
\caption{The behavior of $f(r)$ versus $r$ for $l=1.5$, $b=0.4$, $q=1.4$, $%
z=1.5$ and $m=1.5$.}
\label{fig1}
\end{figure*}

One of the properties of dilaton field is that it couples with gauge fields.
In the presence of dilaton field $\Phi $, the Lagrangian of the logarithmic
electrodynamics get modified as well. In this case, the Lagrangian density
of the logarithmic gauge field coupled to the dilaton field in ($n+1$%
)-dimensions can be written as \cite{SNZ2}%
\begin{equation}
L\left( F,\Phi \right) =-8\beta ^{2}e^{4\lambda \Phi /(n-1)}\ln \left( 1+%
\frac{e^{-8\lambda \Phi /(n-1)}F}{8\beta ^{2}}\right) ,
\end{equation}%
where $\beta $ is the nonlinear parameter and $\lambda $ is a constant. The
large $\beta $ limit of $L\left( F,\Phi \right) $ reproduces the linear
Maxwell electrodynamics coupled to the dilaton field \cite{CHM,Shey3} 
\begin{equation}
L(F,\Phi )=-e^{-4\lambda \Phi /(n-1)}F+\frac{e^{-12\lambda \Phi /(n-1)}F^{2}%
}{16\beta ^{2}}+O\left( \frac{1}{\beta ^{4}}\right) .  \label{larBeta}
\end{equation}%
In this paper, we look for asymptotic Lifshitz topological black hole
solutions. Thus we assume the line elements of the metric is \cite%
{Mann,tario} 
\begin{equation}
ds^{2}=-\frac{r^{2z}}{l^{2z}}f(r)dt^{2}+{\frac{l^{2}}{r^{2}}}\frac{dr^{2}}{%
f(r)}+r^{2}d\Sigma _{k}^{(n-1)},  \label{met}
\end{equation}%
where $z$ is dynamical critical exponent and $k=0,\pm 1$ determines the sign
of constant curvature $(n-1)(n-2)k$ of ($n-1$)-dimensional hypersurface with
the line element $d\Sigma _{k}^{(n-1)}$ and volume $\omega _{n-1}$. In order
to respect the Lifshitz symmetry, we require $f(r)\rightarrow 1$ as $%
r\rightarrow \infty $. We desire to consider the string-generated
Einstein-dilaton model \cite{Polch} with two Maxwell and one logarithmic
gauge fields. The Lagrangian density of this theory in Einstein frame is%
\begin{equation}
\mathcal{L}=\frac{1}{16\pi }\left( \mathcal{R}-\frac{4}{n-1}(\nabla \Phi
)^{2}-2\Lambda +L(F,\Phi )-\sum\limits_{i=1}^{2}e^{-4/(n-1)\lambda _{i}\Phi
}H_{i}\right) ,  \label{Lag}
\end{equation}%
where $\mathcal{R}$ is Ricci scalar and $\Lambda $ and $\lambda _{i}$'s are
some constants. $H_{i}=(H_{i})_{\mu \nu }(H_{i})^{\mu \nu }$ in which $%
(H_{i})_{\mu \nu }=\partial _{\lbrack \mu }(B_{i})_{\nu ]}$ where $%
(B_{i})_{\nu }$ are the gauge potentials. Varying the action $S=\int d^{n+1}x%
\sqrt{-g}\mathcal{L}$ with respect to metric $g_{\mu \nu }$, dilaton field $%
\Phi $ and gauge potentials $A_{\mu }$ and $(B_{i})_{\mu }$'s, one can
derive the corresponding equations of motion as 
\begin{eqnarray}
\mathcal{R}_{\mu \nu } &=&\frac{g_{\mu \nu }}{n-1}\left[ 2\Lambda
+2L_{F}F-L(F,\Phi )-\sum\limits_{i=1}^{2}H_{i}e^{-4\lambda _{i}\Phi /(n-1)}%
\right]   \notag \\
&+&\frac{4}{n-1}\partial _{\mu }\Phi \partial _{\nu }\Phi -2L_{F}F_{\mu
\lambda }F_{\nu }^{\text{ \ }\lambda }+2\sum\limits_{i=1}^{2}e^{-4\lambda
_{i}\Phi /(n-1)}\left( H_{i}\right) _{\mu \lambda }\left( H_{i}\right) _{\nu
}^{\text{ \ }\lambda },  \label{FE1}
\end{eqnarray}%
\begin{eqnarray}
\nabla ^{2}\Phi +\frac{n-1}{8}L_{\Phi }+\sum\limits_{i=1}^{2}\frac{{\lambda
_{i}}}{2}e^{-{4\lambda _{i}\Phi }/({n-1})}H_{i} &=&0,  \label{FE2} \\
\triangledown _{\mu }\left( L_{F}F^{\mu \nu }\right)  &=&0,  \label{FE3} \\
\triangledown _{\mu }\left( e^{-{4\lambda _{i}\Phi }/({n-1})}\left(
H_{i}\right) ^{\mu \nu }\right)  &=&0,  \label{FE4}
\end{eqnarray}%
where $L_{F}=\partial L/\partial F$ and $L_{\Phi }=\partial L/\partial \Phi $%
. Lifshitz black hole solutions of Einstein-dilaton gravity with Maxwell 
\cite{tario}, power Maxwell \cite{Deh2} and exponential nonlinear \cite{KDSD}
electrodynamics have been explored. Clearly, in the limiting case where $%
\beta \rightarrow \infty $, one may expect that our solutions reduce to
linear Maxwell case constructed in \cite{tario} and in \cite{Deh2} when the
power of electrodynamics Lagrangian is equal to $1$. It is notable to
mention that our definitions are so that in the linear limit, our solutions
directly reproduce the results of \cite{Deh2} for $p=1$. First of all, we
solve differential equations (\ref{FE3}) and (\ref{FE4}) by using the metric
(\ref{met}). We find%
\begin{equation}
F_{rt}=\frac{2qe^{4\lambda \Phi /(n-1)}}{(\Upsilon +1)r^{n-z}},  \label{FF}
\end{equation}%
\begin{equation}
\left( H_{i}\right) _{rt}=\frac{q_{i}}{r^{z-n}}e^{4\lambda _{i}\Phi /(n-1)},
\label{HH}
\end{equation}%
where 
\begin{equation}
\Upsilon \equiv \sqrt{1+\frac{q^{2}l^{2z-2}}{\beta ^{2}r^{2n-2}}}
\end{equation}%
and $q$ and $q_{i}$'s are some integration constants. As we will see, $q$ is
related to the electric charge of the black hole. Expanding (\ref{FF}) for $%
\beta \rightarrow \infty $, yields 
\begin{equation}
F_{rt}=\frac{qe^{4\lambda \Phi /(n-1)}}{r^{n-z}}-\frac{q^{3}l^{2z-2}e^{4%
\lambda \Phi /(n-1)}}{4r^{3n-z-2}\beta ^{2}}+O\left( \frac{1}{\beta ^{4}}%
\right) ,  \label{Frt ex}
\end{equation}%
The first term in Eq. (\ref{Frt ex}) is the linear Maxwell one presented in 
\cite{Deh2}. The second term is the leading order nonlinear correction term
to the Maxwell field. Substituting solutions (\ref{FF}) and (\ref{HH}) into
the field equations (\ref{FE1}) and (\ref{FE2}), one arrives at four
differential equations 
\begin{equation}
\frac{(n-1)\left( r^{n}f\right) ^{\prime }}{2l^{2}r^{n-1}}+\frac{2r^{2}f\Phi
^{\prime 2}}{(n-1)l^{2}}+\Lambda -\frac{(n-1)(n-2)k}{2r^{2}}%
+\sum\limits_{i=1}^{2}\frac{q_{i}^{2}e^{4\lambda _{i}\Phi /(n-1)}}{%
l^{2(1-z)}r^{2(n-1)}}+\Xi =0,  \label{Eq1}
\end{equation}%
\begin{equation}
\frac{(n-1)\left( r^{n+2\left( z-1\right) }f\right) ^{\prime }}{%
2l^{2}r^{n+2z-3}}-\frac{2r^{2}f\Phi ^{\prime 2}}{(n-1)l^{2}}+\Lambda -\frac{%
(n-1)(n-2)k}{2r^{2}}+\sum\limits_{i=1}^{2}\frac{q_{i}^{2}e^{4\lambda
_{i}\Phi /(n-1)}}{l^{2(1-z)}r^{2(n-1)}}+\Xi =0,  \label{Eq2}
\end{equation}%
\begin{gather}
\frac{r^{2}f^{\prime \prime }}{2l^{2}}+\frac{(2n+3z-3)rf^{\prime }}{2l^{2}}+%
\frac{2r^{2}f\Phi ^{\prime 2}}{(n-1)l^{2}}+\frac{(2z^{2}+2(n-2)z+(n-1)(n-2))f%
}{2l^{2}}  \notag \\
+\Lambda -\frac{(n-3)(n-2)k}{2r^{2}}-\sum\limits_{i=1}^{2}\frac{%
q_{i}^{2}e^{4\lambda _{i}\Phi /(n-1)}}{l^{2(1-z)}r^{2(n-1)}}-4\beta
^{2}e^{4\lambda \Phi /(n-1)}\ln \left( \frac{\Upsilon +1}{2}\right) =0,
\label{Eq3}
\end{gather}%
\begin{equation}
\frac{\left( r^{n+z}f\Phi ^{\prime }\right) ^{\prime }}{l^{2}r^{n+z-2}}%
-\sum\limits_{i=1}^{2}\frac{q_{i}^{2}\lambda _{i}e^{4\lambda _{i}\Phi /(n-1)}%
}{l^{2(1-z)}r^{2(n-1)}}-\lambda \Xi =0.  \label{Eq4}
\end{equation}%
where $\Phi =\Phi \left( r\right) $ and%
\begin{equation*}
\Xi \equiv 4\beta ^{2}e^{4\lambda \Phi /(n-1)}\left[ \Upsilon -1-\ln (\frac{%
\Upsilon +1}{2})\right] .
\end{equation*}%
Combining (\ref{Eq1}) and (\ref{Eq2}), we find 
\begin{equation}
4r^{2}\Phi ^{\prime 2}=\left( n-1\right) ^{2}\left( z-1\right) ,
\end{equation}%
which has the solution 
\begin{equation}
{\Phi (r)}{=\ln \left( \frac{r}{b}\right) }^{\xi }{,}\text{ \ \ \ \ \ }\xi =%
\frac{(n-1)\sqrt{z-1}}{2},  \label{Phi}
\end{equation}%
where $b$ is an integration constant with dimension of length. Solution (\ref%
{Phi}) implies $z\geq 1$. With $\Phi \left( r\right) $\ at hand, we can
obtain the function $f(r)$\ form Eqs. (\ref{Eq1})-(\ref{Eq4}) as%
\begin{eqnarray}
f(r) &=&-\frac{2l^{2}\Lambda }{\left( n+z-1\right) \left( n+z-2\right) }-%
\frac{m}{r^{n+z-1}}+\frac{(n-2)^{2}kl^{2}}{(n+z-3)^{2}r^{2}}+\frac{8\beta
^{2}l^{2}b^{2z-2}}{(n-1)(n-z+1)r^{2z-2}}  \notag \\
&&\times \left\{ \frac{2n-z}{n-z+1}+\ln \left( \frac{1+\Upsilon }{2}\right) -%
\frac{(n-1)}{(n-z+1)}\mathbf{F}\left( \frac{1}{2},X,X+1,1-\Upsilon
^{2}\right) -\mathbf{F}\left( -\frac{1}{2},X,X+1,1-\Upsilon ^{2}\right)
\right\} ,  \notag \\
&&  \label{F0}
\end{eqnarray}%
where%
\begin{equation*}
X=\frac{z-n-1}{2n-2},
\end{equation*}%
$\mathbf{F}$\ is the hypergeometric function and $m$\ is a constant related
to the mass of the black hole. Using the hypergeometric identities \cite%
{abram}%
\begin{equation}
(y-w-1)\mathbf{F}\left( w,x,y,s\right) +w\mathbf{F}\left( w+1,x,y,s\right)
-\left( y-1\right) \mathbf{F}\left( w,x,y-1,s\right) =0,  \label{hyperid1}
\end{equation}%
and%
\begin{equation}
\mathbf{F}\left( w,x,x,s\right) =\left( 1-s\right) ^{-w},  \label{hyperid2}
\end{equation}%
one can rewrite $f(r)$\ in a more simple form%
\begin{eqnarray}
f(r) &=&-\frac{2l^{2}\Lambda }{\left( n+z-1\right) \left( n+z-2\right) }-%
\frac{m}{r^{n+z-1}}+\frac{(n-2)^{2}kl^{2}}{(n+z-3)^{2}r^{2}}+\frac{8\beta
^{2}l^{2}b^{2z-2}}{(n-1)(n-z+1)r^{2z-2}}  \notag \\
&&\times \left\{ \frac{2n-z}{n-z+1}+\ln \left( \frac{1+\Upsilon }{2}\right) +%
\frac{(n-z+1)\Upsilon }{z-2}-\frac{(n-1)^{2}}{(z-2)(n-z+1)}\mathbf{F}\left( 
\frac{1}{2},\frac{z-n-1}{2n-2},\frac{n+z-3}{2n-2},1-\Upsilon ^{2}\right)
\right\} .  \notag \\
&&  \label{F1}
\end{eqnarray}%
Let us note that although at the first glance relation (\ref{F1}) seems
divergent in $z=2$, $f(r)$\ is not really diverging at this point as one can
see from (\ref{F0}). The above solutions will fully satisfy the system of
Eqs. (\ref{Eq1})-(\ref{Eq4}) provided, 
\begin{gather}
\lambda =-\sqrt{z-1},\text{ \ \ \ \ }\lambda _{1}=\frac{n-1}{\sqrt{z-1}},%
\text{ \ \ \ \ }\lambda _{2}=\frac{n-2}{\sqrt{z-1}},  \notag \\
q_{1}^{2}=-\frac{\Lambda (z-1)b^{2(n-1)}}{(z+n-2)l^{2(z-1)}},\text{ \ \ \ \ }%
q_{2}^{2}=\frac{k(n-1)(n-2)(z-1)b^{2(n-2)}}{2(z+n-3)l^{2(z-1)}}.
\label{Constants}
\end{gather}%
Note that $q$ is hidden in $\Upsilon =\sqrt{1+q^{2}l^{2z-2}/(\beta
^{2}r^{2n-2})}$ in (\ref{F1}). The reality of $q_{2}$ requires that $k\neq -1
$ except for $z=1$. Thus, hereafter, we consider the black branes ($k=0$)
and black holes ($k=1$) in the general cases with $z\neq 1$. Also, reality
of $q_{1}$ implies $\Lambda <0$. As we mentioned above the asymptotic
Lifshitz behavior implies that $f(r)\rightarrow 1$ as $r\rightarrow \infty $%
. However, from Eq. (\ref{F1}) we have 
\begin{equation*}
\lim_{r\rightarrow \infty }f(r)=-\frac{2l^{2}\Lambda }{\left( n+z-1\right)
\left( n+z-2\right) }.
\end{equation*}%
Therefore, in order to have appropriate asymptotic behavior for $f(r)$ we
fix $\Lambda $ as 
\begin{equation}
\Lambda =-\frac{(n+z-1)(n+z-2)}{2l^{2}},  \label{Lambda}
\end{equation}%
which is negative ($\Lambda <0$), as the reality of $q_{1}$ implies. Hence,
the final form of $f(r)$ is 
\begin{eqnarray}
f(r) &=&1-\frac{m}{r^{n+z-1}}+\frac{(n-2)^{2}kl^{2}}{(n+z-3)^{2}r^{2}}+\frac{%
8\beta ^{2}l^{2}b^{2z-2}}{(n-1)(n-z+1)r^{2z-2}}  \notag \\
&&\times \left\{ \frac{2n-z}{n-z+1}+\frac{(n-z+1)\Upsilon }{z-2}+\ln \left( 
\frac{1+\Upsilon }{2}\right) -\frac{(n-1)^{2}}{(z-2)(n-z+1)}\mathbf{F}\left( 
\frac{1}{2},\frac{z-n-1}{2n-2},\frac{n+z-3}{2n-2},1-\Upsilon ^{2}\right)
\right\} .  \notag \\
&&  \label{F}
\end{eqnarray}%
The behavior of $f(r)$ for large $\beta $, may be written 
\begin{eqnarray}
f(r) &=&1-\frac{m}{r^{n+z-1}}+\frac{(n-2)^{2}kl^{2}}{(n+z-3)^{2}r^{2}}+\frac{%
2q^{2}b^{2z-2}l^{2z}}{\left( n-1\right) \left( n+z-3\right) r^{2n+2z-4}}-%
\frac{q^{4}b^{2z-2}l^{4z-2}}{4\left( n-1\right) \left( 3n+z-5\right) \beta
^{2}r^{4n+2z-6}}+O\left( \frac{1}{\beta ^{4}}\right) .  \notag \\
&&  \label{expf}
\end{eqnarray}%
When $\beta \rightarrow \infty $, this solution recovers the Lifshitz black
holes in Einstein-Maxwell-dilaton gravity \cite{tario,Deh2}, as expected.
Figure \ref{fig1} shows the behavior of $f(r)$ for different values of $%
\beta $ correspond to Lifshitz black branes ($k=0$) and black holes ($k=1$).
This figure exhibits that it is possible to have black solutions with one or
two horizons. We can calculate the Hawking temperature of outermost horizon $%
r_{+}$ as

\begin{equation}
T=\frac{r_{+}^{z+1}f^{\prime }\left( r_{+}\right) }{4\pi l^{z+1}}=\frac{%
(n+z-1)r_{+}^{z}}{4\pi l^{z+1}}+\allowbreak \frac{(n-2)^{2}kr_{+}^{z-2}}{%
4\pi (n+z-3)l^{z-1}}+\frac{2\beta ^{2}l^{1-z}b^{2z-2}}{\pi (n-1)r_{+}^{z-2}}%
\left[ 1-\Upsilon _{+}+\ln \left( \frac{\Upsilon _{+}+1}{2}\right) \right] .
\label{Temp}
\end{equation}%
where $\Upsilon _{+}=\Upsilon (r_{+})$. In the next section we shall study
thermodynamics of Lifshitz black branes/holes we obtained in this section. 

\section{THERMODYNAMICS OF LIFSHITZ SOLUTIONS \label{Therm}}

In this section, we want to study thermodynamics of Lifshitz black
holes/branes. The temperature of our solutions on the horizon was calculated
in previous section. In order to find the entropy of these Lifshitz
solutions we can use the so-called area law which states that the entropy of
the black hole is quarter of the event horizon area \cite{Beck}. The entropy
of almost all kinds of black holes in Einstein gravity including dilaton
ones is computed by using this near universal law \cite{hunt}. Hence, the
entropy of the obtained Lifshitz solutions per unit volume $\omega _{n-1}$
can be calculated as 
\begin{equation}
S=\frac{r_{+}^{n-1}}{4}.  \label{entropy}
\end{equation}%
Now, we turn to calculation of electric charge of Lifshitz black holes. We
use the nonlinear logarithmic electrodynamics. The well-known Gauss law for
this nonlinear electrodynamics can be given by 
\begin{equation}
Q=\frac{\,{1}}{4\pi }\int r^{n-1}L_{F}F_{\mu \nu }n^{\mu }u^{\nu }d{\Sigma },
\label{chdef}
\end{equation}%
where $u^{\nu }$ and $u^{\mu }$ are the unite timelike and spacelike normals
to a sphere of radius $r$ given as 
\begin{equation}
n^{\mu }=\frac{1}{\sqrt{-g_{tt}}}dt=\frac{l^{z}}{r^{z}\sqrt{f(r)}}dt,\text{
\ \ \ \ }u^{\nu }=\frac{1}{\sqrt{g_{rr}}}dr=\frac{r\sqrt{f(r)}}{l}dr.
\end{equation}%
Therefore, the electric charge per unit volume $\omega _{n-1}$ is obtained
as 
\begin{equation}
Q=\frac{ql^{z-1}}{4\pi }.  \label{charge}
\end{equation}%
Another conserved quantity of our solutions is mass. We can obtain this
conserved quantity by applying the modified subtraction method of Brown and
York \cite{modBY}. Thus, the mass per unit volume is computed as (see Ref. 
\cite{Deh2} for more details) 
\begin{equation}
M=\frac{(n-1)m}{16\pi l^{z+1}},  \label{Mass}
\end{equation}%
where $m$ can be calculated by using this fact that $f(r_{+})=0$. Therefore,
one obtains 
\begin{eqnarray}
m &=&r_{+}^{n+z-1}+\frac{(n-2)^{2}kl^{2}r_{+}^{n+z-3}}{(n+z-3)^{2}}+\frac{%
8\beta ^{2}l^{2}b^{2z-2}r_{+}^{n-z+1}}{(n-1)(n-z+1)}  \notag \\
&&\times \left\{ \frac{2n-z}{n-z+1}+\frac{(n-z+1)\Upsilon _{+}}{z-2}+\ln
\left( \frac{1+\Upsilon _{+}}{2}\right) -\frac{(n-1)^{2}}{(z-2)(n-z+1)}%
\mathbf{F}\left( \frac{1}{2},\frac{z-n-1}{2n-2},\frac{n+z-3}{2n-2}%
,1-\Upsilon _{+}^{2}\right) \right\} .  \notag \\
&&  \label{massmet}
\end{eqnarray}%
Since from one side $m=m(r_{+},q)$ (note that $q$ is hidden in $\Upsilon
_{+} $) and from another side $r_{+}$ and $q$ are related to entropy and
charge through (\ref{entropy}) and (\ref{charge}), respectively, one can
re-express $m$ and consequently $M$ in terms of extensive quantities $S$ and 
$Q$. The desired Smarr formula $M(S,Q)$ is therefore 
\begin{eqnarray}
M\left( S,Q\right) &=&\frac{(n-1)\left( 4S\right) ^{(n+z-1)/(n-1)}}{16\pi
l^{z+1}}+\frac{\left( n-1\right) (n-2)^{2}k(4S)^{(n+z-3)/(n-1)}}{16\pi
l^{z-1}\left( n+z-3\right) ^{2}}+\frac{\beta ^{2}b^{2z-2}\left( 4S\right)
^{\left( n-z+1\right) /\left( n-1\right) }}{2\pi (n-z+1)l^{z-1}}  \notag \\
&&\times \left\{ \frac{2n-z}{n-z+1}+\frac{(n-z+1)\Gamma }{z-2}+\ln \left( 
\frac{1+\Gamma }{2}\right) -\frac{(n-1)^{2}}{(z-2)(n-z+1)}\mathbf{F}\left( 
\frac{1}{2},\frac{z-n-1}{2n-2},\frac{n+z-3}{2n-2},1-\Gamma ^{2}\right)
\right\} ,  \notag \\
&&
\end{eqnarray}%
where $\Gamma =\sqrt{1+\pi ^{2}Q^{2}/(\beta ^{2}S^{2})}$. One can expand $%
M(S,Q)$ for large values of $\beta $ to arrive at 
\begin{eqnarray}
M(S,Q) &=&\frac{(n-1)(4S)^{(n+z-1)/(n-1)}}{16\pi l^{z+1}}+\frac{%
(n-1)(n-2)^{2}k(4S)^{(n+z-3)/(n-1)}}{16\pi (n+z-3)^{2}l^{z-1}}+\frac{2\pi
Q^{2}b^{2z-2}(4S)^{(3-n-z)/(n-1)}}{(n+z-3)l^{z-1}}  \notag \\
&&-\frac{16\pi ^{3}Q^{4}b^{2z-2}(4S)^{(5-3n-z)/(n-1)}}{4(3n+z-5)l^{z-1}\beta
^{2}}+O\left( \frac{1}{\beta ^{4}}\right) ,
\end{eqnarray}%
which is the Smarr-type formula obtained for the Lifshitz black holes of EMd
theory in the limit of $\beta \rightarrow \infty $ \cite{Deh2}. We also
introduce the conjugate intensive quantities corresponding to entropy and
charge namely temperature and electric potential as 
\begin{equation}
T=\left( \frac{\partial M}{\partial S}\right) _{Q}\text{ \ \ \ \ and \ \ \ \ 
}U=\left( \frac{\partial M}{\partial Q}\right) _{S}.  \label{intqua}
\end{equation}%
The electric potential $U$, measured at infinity with respect to the horizon 
$r_{+}$, is principally defined as 
\begin{equation}
U=A_{\mu }\chi ^{\mu }\left\vert _{r\rightarrow \infty }-A_{\mu }\chi ^{\mu
}\right\vert _{r=r_{+}},  \label{Pot}
\end{equation}%
where $\chi =\partial _{t}$ is the null generator of the horizon. In order
to find electric potential $U$, we first have to calculate the gauge
potential $A_{t}$. The gauge potential $A_{t}$ corresponding to the
electromagnetic field (\ref{FF}) is given by $A_{t}(r)=\int F_{rt}dr$. It is
a matter of calculations to show that 
\begin{equation}
A_{t}=\mu +\frac{2\beta ^{2}b^{2z-2}l^{2-2z}r^{n-z+1}}{(n-z+1)q}(\Upsilon
-1)-\frac{2q(n-1)b^{2z-2}}{(1+n-z)(n+z-3)r^{n+z-3}}\mathbf{F}\left( \frac{1}{%
2},\frac{z+n-3}{2n-2},\frac{n+z-5}{n-1},1-\Upsilon ^{2}\right) .
\label{Apot}
\end{equation}%
One can check that $A_{t}$ reduces to finite value $\mu $ at infinity.
Requiring the fact that $A_{t}\left( r_{+}\right) =0$, one gets 
\begin{equation}
\mu =-\frac{2\beta ^{2}b^{2z-2}l^{2-2z}r_{+}^{n-z+1}}{(n-z+1)q}(\Upsilon
_{+}-1)+\frac{2q(n-1)b^{2z-2}}{(1+n-z)(n+z-3)r_{+}^{n+z-3}}\mathbf{F}\left( 
\frac{1}{2},\frac{z+n-3}{2n-2},\frac{n+z-5}{n-1},1-\Upsilon _{+}^{2}\right) ,
\end{equation}%
Note that $\mu $ is commonly referred to as chemical potential of the
thermodynamical system lives on boundary. Using Eqs. (\ref{Pot}) and (\ref%
{Apot}) the electric potential may be obtained as 
\begin{equation}
U=\mu .  \label{Poten}
\end{equation}%
If we consider $S$ and $Q$ as a complete set of extensive quantities for $%
M(S,Q)$, it is confirmed numerically that the intensive quantities
corresponding to $S$ and $Q$ namely temperature $T$ and electric potential $%
U $, coincide with Eqs. (\ref{Temp}) and (\ref{Poten}), respectively. Thus,
the first law of thermodynamics 
\begin{equation}
dM=TdS+UdQ.  \label{TFL}
\end{equation}%
is satisfied for our obtained Lifshitz black branes/holes. In the remaining
part of this paper, we turn to study the holographic conductivity of a ($2+1$%
)-dimensional system lives on the boundary of brane of a four dimensional
bulk.

\section{Holographic Electrical Conductivity \label{Cond}}

In this section, we will focus on studying gauge/gravity duality for
Lifshitz black brane solutions. In particular, we obtain the AC conductivity
as a function of frequency for a ($2+1$)-dimensional system lives on the
boundary of brane. In order to make the effects of nonlinearity on the
conductivity more clear, we first review the calculation of this quantity
for the linear Maxwell case \cite{wu1,wu2}. Then, we turn to the case with
nonlinear logarithmic electrodynamics. In what follows, we set $l=b=r_{+}=1$%
. We take the planar ($3+1$)-dimensional metric for the bulk as 
\begin{equation}
ds^{2}=-\mathcal{F}(u)u^{-2z}dt^{2}+\left[ \mathcal{F}(u)u^{2}\right] ^{-1}{%
du^{2}}+u^{-2}(dx^{2}+dy^{2}),  \label{metcond}
\end{equation}%
which can be obtained from (\ref{met}) by defining $u=1/r$. Therefore, the
black brane horizon sits at $u=1$ and the three-dimensional system lives at $%
u=0$ (brane boundary). For linear Maxwell case, $\mathcal{F}(u)$ in (\ref%
{metcond}) is calculated by substituting $u=1/r$, $n=3$ and $k=0$ in (\ref%
{expf}) and taking the $\beta \rightarrow \infty $ limit. Thus, we get 
\begin{equation}
\mathcal{F}(u)=1-mu^{z+2}+q^{2}z^{-1}u^{2z+2}.
\end{equation}%
Now, we perturb the vector potential and the metric by turning on $%
A_{x}(u)e^{-i\omega t}$ and $g_{tx}\left( u\right) e^{-i\omega t}$ and
arrive at two additional equations as 
\begin{equation}
A_{x}^{\prime \prime }+\left[ \mathcal{F}^{\prime }\mathcal{F}%
^{-1}+3(1-z)u^{-1}\right] A_{x}^{\prime }+u^{2z-2}\mathcal{F}^{-2}\left[
\omega ^{2}A_{x}-qu^{z}\mathcal{F}\left( 2g_{tx}+ug_{tx}^{\prime }\right) %
\right] =0,  \label{EMaxPert}
\end{equation}%
and%
\begin{equation}
2g_{tx}+ug_{tx}^{\prime }=4qu^{2-z}A_{x},  \label{Einpert}
\end{equation}%
where the prime indicates the derivative with respect to $u$. Eliminating $%
g_{tx}$ from Eq. (\ref{EMaxPert}) through (\ref{Einpert}), we can easily
obtain a linearized equation for the gauge field $A_{x}$%
\begin{equation}
A_{x}^{\prime \prime }+\left[ \mathcal{F}^{\prime }\mathcal{F}%
^{-1}+3(1-z)u^{-1}\right] A_{x}^{\prime }+u^{2z-2}\mathcal{F}^{-2}\left[
\omega ^{2}-4q^{2}u^{2}\mathcal{F}\right] A_{x}=0.  \label{Axeq}
\end{equation}%
From gauge/gravity duality, we know that the expectation value of current is
given by \cite{hart}%
\begin{equation}
\left\langle J_{x}\right\rangle =\left. \frac{\partial \mathfrak{L}}{%
\partial \left( \partial _{u}\delta A_{x}\right) }\right\vert _{u=0},
\label{Jx}
\end{equation}%
where $\mathfrak{L}=\sqrt{-g}\mathcal{L}$ in which $\mathcal{L}$ was
introduced in (\ref{Lag}) and $\delta A_{x}=A_{x}e^{-i\omega t}$. Therefore,
we can calculate conductivity through Ohm's law as%
\begin{equation}
\sigma \left( \omega \right) =\frac{\left\langle J_{x}\right\rangle }{E_{x}}%
=-\frac{\left\langle J_{x}\right\rangle }{\partial _{t}\delta A_{x}}=-\frac{%
i\left\langle J_{x}\right\rangle }{\omega \delta A_{x}}.  \label{cond}
\end{equation}%
In order to compute conductivity $\sigma \left( \omega \right) $, we need to
know the asymptotic behavior of the perturbative field $A_{x}$ governed by (%
\ref{Axeq}) near the boundary $u=0$. This reads%
\begin{equation}
A_{x}^{\prime \prime }-3(z-1)u^{-1}A_{x}^{\prime }+\omega
^{2}u^{2(z-1)}A_{x}+\cdots =0,
\end{equation}%
which has the solutions

\begin{equation}
A_{x}\left( u\right) =\left\{ 
\begin{array}{cc}
A^{0}+\frac{A^{0}\omega ^{2}}{2z\left( z-2\right) }u^{2z}+A^{1}u^{3z-2}+%
\cdots & \text{for }z\neq 2 \\ 
&  \\ 
A^{0}-\frac{A^{0}\omega ^{2}}{4}\ln \left( u\right) u^{4}+A^{1}u^{4}+\cdots
& \text{for }z=2%
\end{array}%
\right. ,  \label{Axlar}
\end{equation}%
where $A^{0}$ and $A^{1}$ are two constants. Thus, the conductivity for the
linear Maxwell electrodynamics is obtained as \cite{wu1,wu2}%
\begin{equation}
\sigma =\left\{ 
\begin{array}{cc}
\frac{\left( 3z-2\right) A^{1}}{4\pi i\omega A^{0}} & \text{for }z\neq 2 \\ 
&  \\ 
\frac{16A^{1}-A^{0}\omega ^{2}}{16\pi i\omega A^{0}} & \text{for }z=2%
\end{array}%
\right. ,  \label{cond1}
\end{equation}%
It is remarkable to note that for $z\geq 2$, there is a divergence term in $%
\mathfrak{L}$\ when we use (\ref{Jx}) and (\ref{cond}) to calculate
conductivity, $\sigma $. However, these terms do not effect on the value of
conductivity and can be easily eliminated by using holographic
re-normalization approach \cite{wu1,wu2,renorm}. In this method, the
divergence is canceled by adding appropriate counterterms to the action. In 
\cite{wu2}, conductivity has been studied for ($3+1$)-dimensional black
branes in the presence of linear Maxwell electrodynamics where $1\leq z\leq
2 $. Here, in (\ref{cond1}), we generalize those results to $z>2$.

Now, we turn to calculate the conductivity on the boundary of the Lifshitz
black branes when the bulk gauge field is in the form of logarithmic
nonlinear electrodynamics. The motivation is to disclose the effects of
nonlinearity on the holographic conductivity in comparison with linear
Maxwell case. In this case, by transforming $r\rightarrow u=1/r$ in Eq. (\ref%
{F}), $\mathcal{F}(u)$ can be rewritten as%
\begin{equation}
\mathcal{F}(u)=1-mu^{z+2}+\frac{4\beta ^{2}u^{2z-2}}{(4-z)}\left\{ \frac{6-z%
}{4-z}+\frac{(4-z)\Upsilon _{u}}{z-2}+\ln \left( \frac{1+\Upsilon _{u}}{2}%
\right) -\frac{4}{(z-2)(4-z)}\mathbf{F}\left( \frac{1}{2},\frac{z-4}{4},%
\frac{z}{4},1-\Upsilon _{u}^{2}\right) \right\} ,
\end{equation}%
where $\Upsilon _{u}=\sqrt{1+q^{2}u^{4}/\beta ^{2}}$. Perturbative equations
of motion coming from turning on $A_{x}(u)e^{-i\omega t}$ and $g_{tx}\left(
u\right) e^{-i\omega t}$ in the bulk for nonlinear electrodynamics are%
\begin{equation}
A_{x}^{\prime \prime }+\left[ \frac{3(1-z)}{u}+\frac{\mathcal{F}^{\prime }}{%
\mathcal{F}}+\frac{4q^{2}u^{3}}{q^{2}u^{4}+\beta ^{2}(1+\Upsilon _{u})^{2}}%
\right] A_{x}^{\prime }+\frac{\omega ^{2}u^{2z-2}}{\mathcal{F}^{2}}A_{x}=%
\frac{2qu^{3z-2}}{(1+\Upsilon _{u})\mathcal{F}}\left(
2g_{tx}+ug_{tx}^{\prime }\right) ,  \label{Eper}
\end{equation}%
and 
\begin{equation}
2g_{tx}+ug_{tx}^{\prime }=4qu^{2-z}A_{x},
\end{equation}%
which give rise to the decoupled equation for the gauge field $A_{x}$%
\begin{equation}
A_{x}^{\prime \prime }+\left[ \frac{3(1-z)}{u}+\frac{\mathcal{F}^{\prime }}{%
\mathcal{F}}+\frac{4q^{2}u^{3}l^{2z-2}}{q^{2}u^{4}+\beta ^{2}(1+\Upsilon
_{u})^{2}}\right] A_{x}^{\prime }+\frac{u^{2z-2}}{\mathcal{F}^{2}}A_{x}\left[
\omega ^{2}-\frac{8q^{2}u^{2}\mathcal{F}}{(1+\Upsilon _{u})}\right] =0.
\label{EEper}
\end{equation}%
One can check that the general behavior of Eq. (\ref{EEper}) near the
boundary $u=0$ is 
\begin{equation}
A_{x}^{\prime \prime }-3(z-1)u^{-1}A_{x}^{\prime }+\omega
^{2}u^{2(z-1)}A_{x}+\cdots =0,
\end{equation}%
which cause the same behavior for the gauge potential near the boundary as (%
\ref{Axlar}). Now, we can compute conductivity in the presence of
logarithmic electrodynamics. Using Eq. (\ref{cond}), we arrive at 
\begin{equation}
\sigma =\left\{ 
\begin{array}{cc}
\frac{\left( 3z-2\right) A^{1}}{4\pi i\omega A^{0}}\left. \left[ 1+\left( 
\frac{\omega A^{0}u^{3-z}}{2\beta }\right) ^{2}\right] ^{-1}\right\vert
_{u=0} & \text{for }z\neq 2, \\ 
&  \\ 
\frac{16A^{1}-A^{0}\omega ^{2}}{16\pi i\omega A^{0}}\left. \left[ 1+\left( 
\frac{\omega A^{0}u}{2\beta }\right) ^{2}\right] ^{-1}\right\vert _{u=0} & 
\text{for }z=2.%
\end{array}%
\right.   \label{holcond}
\end{equation}%
Consequently, the conductivity $\sigma $ in this case for different ranges
of dynamical critical exponent $z$ is%
\begin{equation}
\sigma =\left\{ 
\begin{array}{ll}
\frac{(3z-2)A^{1}}{4\pi i\omega A^{0}} & \text{for }z<3\text{ (}\neq 2\text{)%
}, \\ 
&  \\ 
\frac{16A^{1}-A^{0}\omega ^{2}}{16\pi i\omega A^{0}} & \text{for }z=2, \\ 
&  \\ 
\frac{(3z-2)A^{1}}{4\pi i\omega A^{0}}\left[ 1+\left( \frac{\omega A^{0}}{%
2\beta }\right) ^{2}\right] ^{-1} & \text{for }z=3, \\ 
&  \\ 
0 & \text{for }z>3.%
\end{array}%
\right.   \label{holcond2}
\end{equation}%
It is notable to mention that the same comments as linear Maxwell case given
in the end of previous paragraph about the diverging terms are also valid
here. The above result indicates that for $z<3$, the conductivity has the
same expression as linear Maxwell field. For $z=3$ the conductivity get
modified due to the nonlinear parameter $\beta $ and reduces to the Maxwell
case as $\beta \rightarrow \infty $. However, in contrast to the linear
Maxwell field, the conductivity is zero for $z>3$. The media show the
behavior like the latter case ($z>3$) in which $\sigma =0$\ are known as
"lossless" since conductivity represents power loss within a medium. In such
a media which is called also as "perfect dielectrics", $J=0$ regardless of
the electric field $E$. This means that the electric field $E$\ cannot move
the charge carriers. This result shows that, for $z>3$  the media in gauge
side of gauge/gravity duality represent lossless behavior if the nonlinear
electrodynamics is employed.

In order to have further understanding on the behavior of the conductivity,
we depict the conductivity in terms of the frequency by solving Eq. (\ref%
{EEper}), numerically. For this purpose, we need initial conditions. Let us
look at the solution for the gauge potential $A_{x}$ near the horizon $r_{+}$%
. In order to take into account the causal behavior, we solve Eq. (\ref%
{EEper}) for $A_{x}$ near horizon by performing ingoing wave boundary
condition. Therefore, we receive%
\begin{equation}
A_{x}\left( u\right) =\mathcal{F}(u)^{-i4\pi \omega /T}\Psi (u),  \label{Ax}
\end{equation}%
where $T$ is temperature and%
\begin{equation}
\Psi (u)=1+a(u-1)+b(u-1)^{2}+\cdots .
\end{equation}%
where $a,b,\ldots $ are some constants to be determined numerically and
considered as initial conditions required for solving differential equation (%
\ref{EEper}), numerically. Substituting Eq. (\ref{Ax}) into Eq. (\ref{EEper}%
), one can obtain the differential equation for $\Psi $. Therefore, $%
a,b,\ldots $can be found by looking for Taylor series expansions of Eq. (\ref%
{EEper}) near the horizon $r_{+}$. With these initial conditions, we are
able to plot the behavior of the conductivity in terms of frequency.

\begin{figure*}[t]
\centering{%
\subfigure[]{
   \label{fig2a}\includegraphics[width=.46\textwidth]{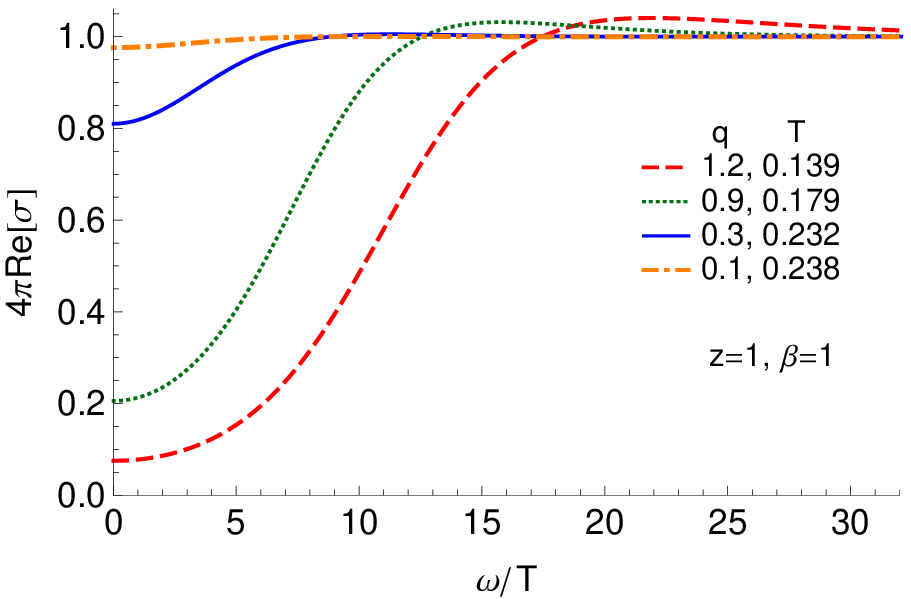}\qquad}} 
\subfigure[]{
   \label{fig2b}\includegraphics[width=.46\textwidth]{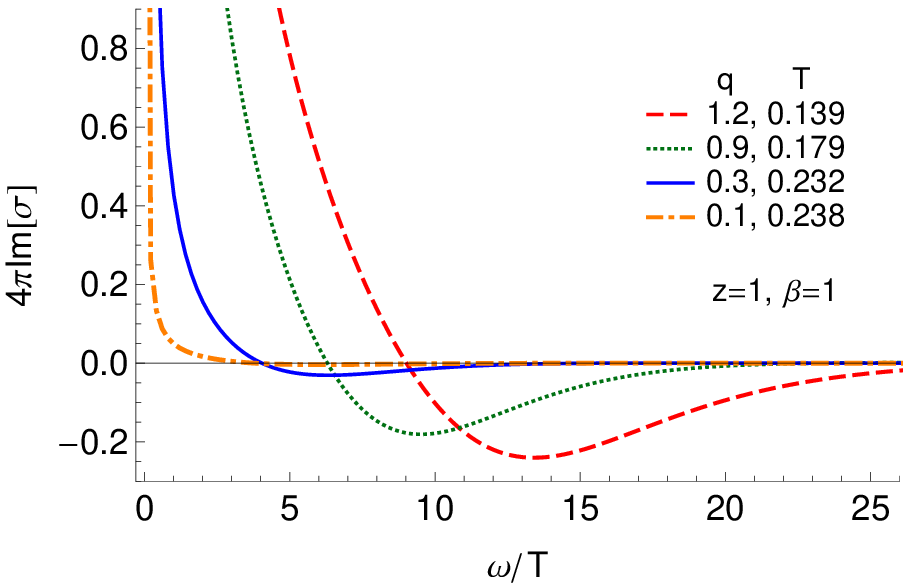}}
\caption{The behaviors of real and imaginary parts of electrical
conductivity $\protect\sigma $ versus $\protect\omega /T$ for $z=1$, $%
\protect\beta =1$ and different values of $q$ with $l=b=r_{+}=1$.}
\label{fig2}
\end{figure*}

\begin{figure*}[t]
\centering{%
\subfigure[]{
   \label{fig3a}\includegraphics[width=.46\textwidth]{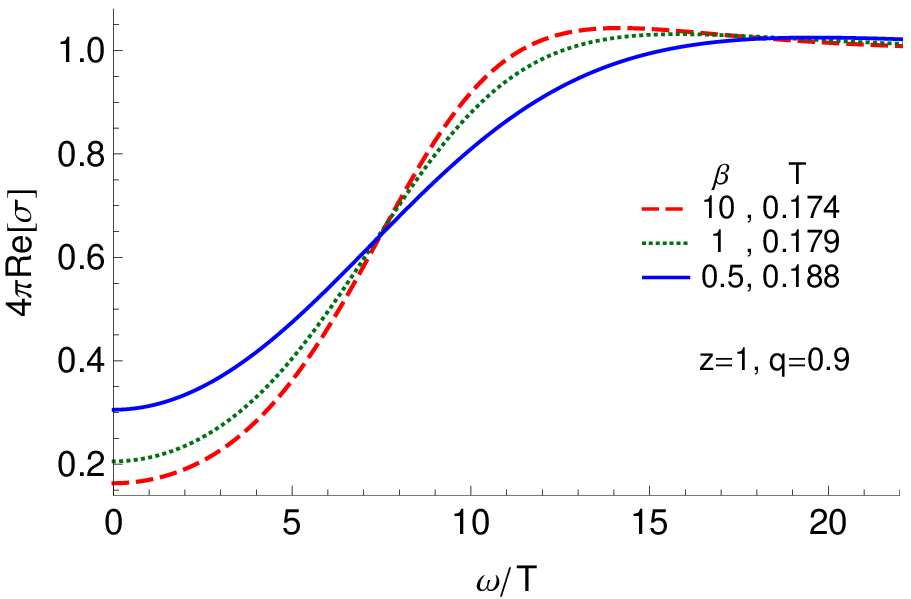}\qquad}} 
\subfigure[]{
   \label{fig3b}\includegraphics[width=.46\textwidth]{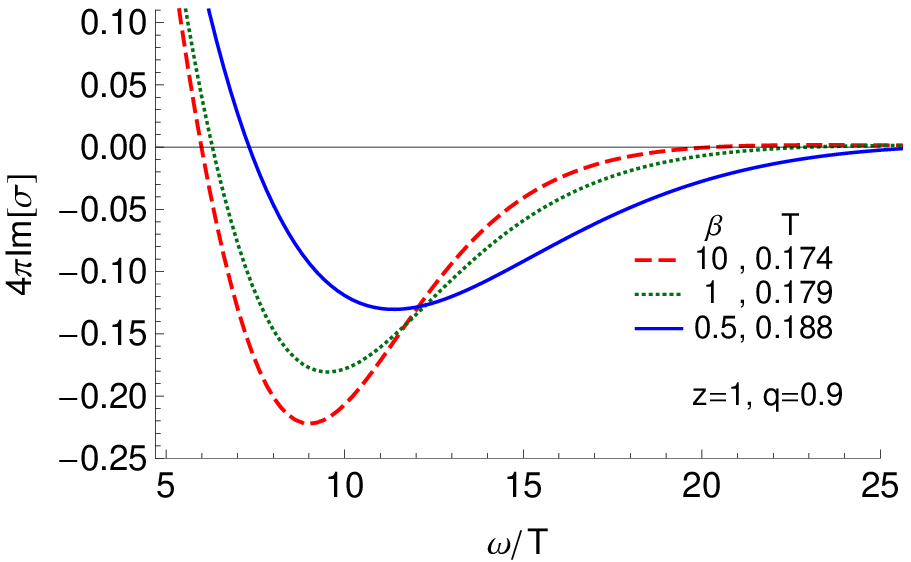}}
\caption{The behaviors of real and imaginary parts of electrical
conductivity $\protect\sigma $ versus $\protect\omega /T$ for $z=1$, $q=0.9$
and different values of $\protect\beta $ with $l=b=r_{+}=1$.}
\label{fig3}
\end{figure*}

Figs. \ref{fig2} and \ref{fig3} show the behavior of conductivity $\sigma $
with respect to frequency per temperature, $\omega /T$, for asymptotic AdS
case ($z=1$). In Fig. \ref{fig2}, this behavior is depicted for different
values of $q$. From Fig. \ref{fig2a} we see that $\sigma _{DC}=$\textrm{Re}$%
[\sigma \left( 0\right) ]$ decreases as $q$ (temperature $T$) increases
(decreases). This figure also shows that the effects of increasing of
frequency is more for larger values of $q$ (lower temperatures).
Furthermore, inspite of different values of $q$ ($T$) there is an asymptotic
value for $\mathrm{Re}[\sigma ]$ in large frequencies. Such behavior for
conductivity has been reported in \cite{exp1} for a graphene system. It is
important to note that the behaviors of real and imaginary parts of
conductivity are not independent and is related to each other via
Kramers-Kronig relations. Fig. \ref{fig3} illustrates the behavior of
conductivity versus frequency for different values of the nonlinear
parameter $\beta $. This figure again confirms that $\sigma _{DC}$ decreases
with decreasing the temperature. As one can see from Fig. \ref{fig3a}, there
is a specific $\omega /T$ (between $5$ and $10$) that inspite of different
values of $\beta $, the conductivity is the same for it.

\begin{figure*}[t]
\centering{%
\subfigure[]{
   \label{fig4a}\includegraphics[width=.46\textwidth]{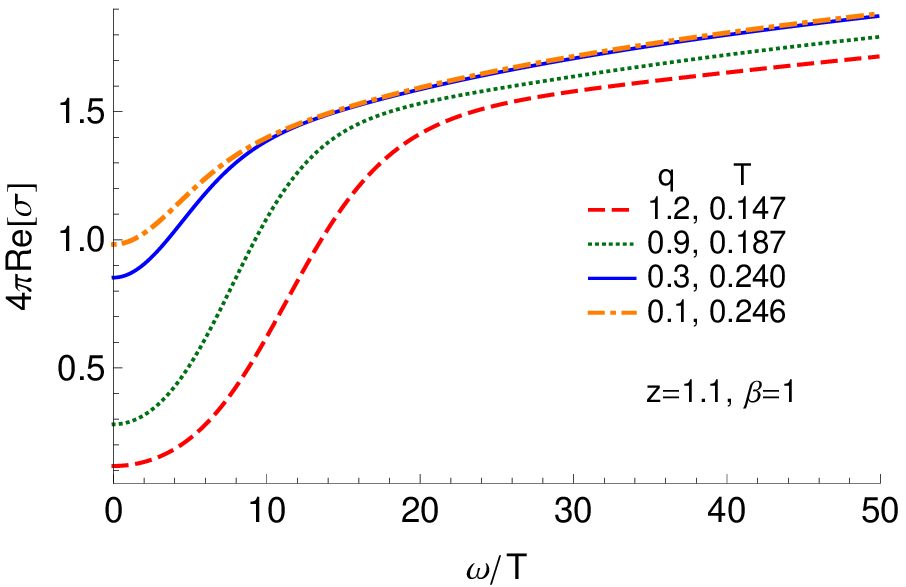}\qquad}} 
\subfigure[]{
   \label{fig4b}\includegraphics[width=.46\textwidth]{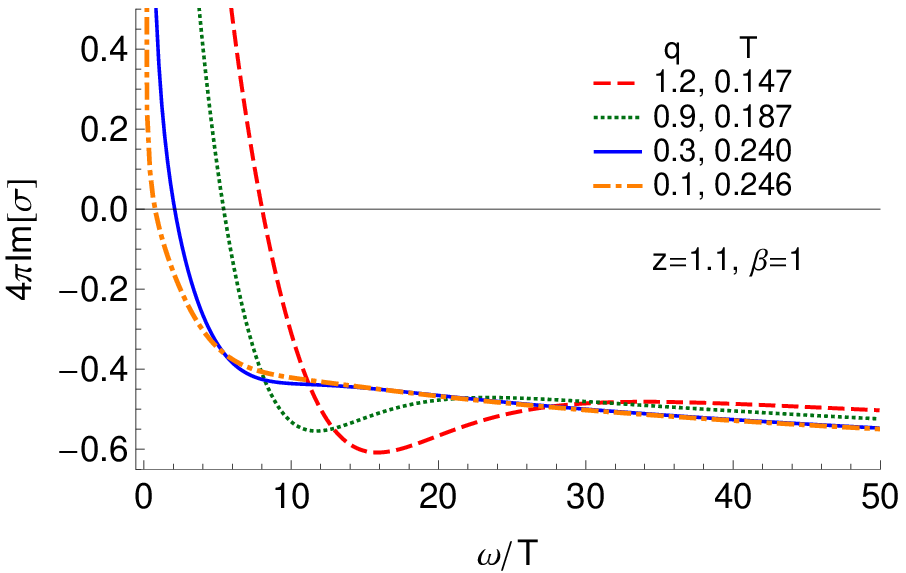}}
\caption{The behaviors of real and imaginary parts of electrical
conductivity $\protect\sigma $ versus $\protect\omega /T$ for $z=1.1$, $%
\protect\beta =1$ and different values of $q$ with $l=b=r_{+}=1$.}
\label{fig4}
\end{figure*}

\begin{figure*}[t]
\centering{%
\subfigure[]{
   \label{fig5a}\includegraphics[width=.46\textwidth]{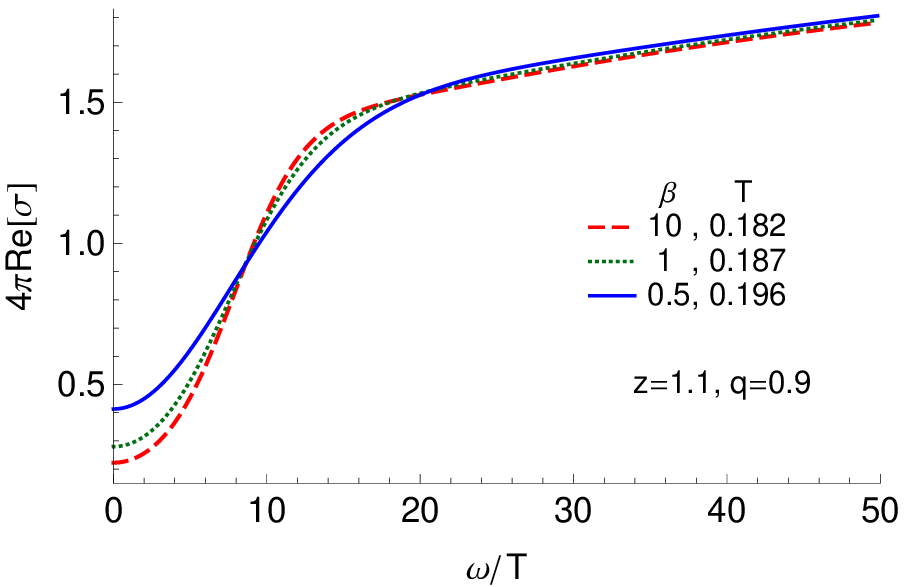}\qquad}} 
\subfigure[]{
   \label{fig5b}\includegraphics[width=.46\textwidth]{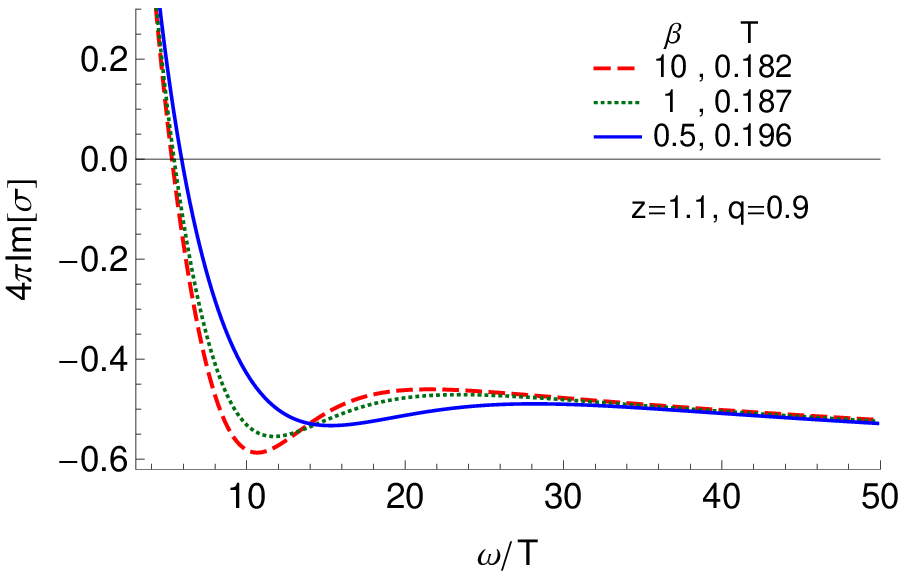}}
\caption{The behaviors of real and imaginary parts of electrical
conductivity $\protect\sigma $ versus $\protect\omega /T$ for $z=1.1$, $%
q=0.9 $ and different values of $\protect\beta $ with $l=b=r_{+}=1$.}
\label{fig5}
\end{figure*}

In Figs. \ref{fig4} and \ref{fig5}, previous cases are illustrated for a
system with Schrodinger-like symmetry, namely $z=1.1$. For small
frequencies, $\sigma _{DC}$ has the same behavior as previous case i.e., it
increases as temperature does. However, the behavior of conductivity for
large frequencies is different. In fact, real part of conductivity has a
positive slope with respect to frequency for large ones. This behavior may
be interpreted as existence of an additional charge carrier rather than the
previous case. This is due to existence of dilaton scalar field in model in
comparison with asymptotic AdS case. Similar behavior for optical
conductivity of single-layer graphene induced by mild oxygen plasma exposure
has been reported in \cite{exp2}. Fig. \ref{fig4} shows that we have more
near behavior for conductivity for smaller $q$'s (higher temperatures).
Finally, from Fig. \ref{fig5}, we see that for a specific value of $\omega
/T $ (near $10$)\textbf{, }the conductivity has the same behavior,
independent of the nonlinear parameter $\beta $. This occurs for isotropic
symmetrical systems ($z=1$) too (see Fig. \ref{fig3a}). 

\section{CLOSING REMARKS}

To summarize, we have constructed a new class of $(n+1)$-dimensional
Lishshitz dilaton black holes/branes in the presence of logarithmic
nonlinear electrodynamics. The obtained solutions in this paper obey the
scaling symmetry $t\rightarrow \lambda ^{z}t$ and $x^{i}\rightarrow \lambda
x^{i}$, comes from the generalized gauge/gravity duality, at $r$-infinity
boundary. We found that the horizon of these spacetime can be an $(n-1)$%
-dimensional hypersurface with positive $(k=1)$ or zero $(k=0)$, constant
curvature. Therefore, our solutions rule out the case with negative
curvature ($k=-1$). We studied thermodynamics of Lifshitz dilaton black
holes/branes and calculated the temperature, entropy, charge, electric
potential and mass of the spacetime. We have also confirmed that these
conserved and thermodynamic quantities satisfy the first law of
thermodynamics on the horizon. This is a necessary requirement for a system
in which one can apply gauge/gravity duality.

Then, we investigated the gauge/gravity duality of Lifshitz black branes by
calculating the holographic conductivity as a function of frequency for a ($%
2+1$)-dimensional system lives on the boundary of a $(3+1)$-dimensional
bulk. First, we reviewed the calculations of the conductivity on the
boundary of Einstein-Maxwell-dilaton black branes and found that it holds
for all values of the dynamical exponent $z$. In this part, we generalized
the study of \cite{wu2} to $z>2$. Then, we extended our study to the case
with logarithmic nonlinear electrodynamics. We found that for $z<3$, the
conductivity has the same expression as linear Maxwell field. For $z=3$ the
conductivity get modified due to the nonlinear parameter $\beta $ as given
in Eq. (\ref{holcond2}), and reduces to the Maxwell case as $\beta
\rightarrow \infty $. However, in contrast to the linear Maxwell case, in
the presence of logarithmic nonlinear gauge field the conductivity is zero
for $z>3$. Taking suitable initial conditions, we have plotted the behavior
of the conductivity in terms of frequency per temperature. We found that for
asymptotic AdS case ($z=1$), $\sigma _{DC}=$\textrm{\ Re}$[\sigma \left(
0\right) ]$ decreases as $q$ (temperature $T$) increases (decreases). Latter
behavior holds for asymptotic Lifshitz solutions ($z>1$). However, the
behavior of conductivity for large frequencies is different. In fact, real
part of conductivity has a positive slope with respect to large frequency
for asymptotic Lifshitz solutions whereas it tends to a constant for
asymptotic AdS ones. This behavior may be interpreted as existence of an
additional charge carrier for systems respecting Lifshitz symmetry rather
than the AdS case. This is due to existence of dilaton scalar field in our
model, comparing with asymptotic AdS case. Similar behavior for optical
conductivity of single-layer graphene induced by mild oxygen plasma exposure
has been reported in \cite{exp2}. Finally, we observed that for a specific
value of $\omega /T$, the conductivity has the same value, independent of
the nonlinear parameter $\beta $, for both asymptotic Lifshitz and
asymptotic AdS solutions.

\acknowledgments{We are grateful to the referee for constructive
and helpful comments which helped us improve the paper
significantly. We also thank the research council of Shiraz
University. This work has been financially supported by the
Research Institute for Astronomy \& Astrophysics of Maragha
(RIAAM), Iran.}

\end{document}